\magnification=1200
\noindent {\bf Duality and Lorentzian Kac-Moody Algebras}
\vskip 10pt
\noindent David I Olive, University of Wales Swansea.
\vskip10pt 
\noindent Invited talk at the Workshop on {\bf Integrable Theories, Solitons and Duality} (1-6 July 2002)
\vskip10pt

I would like to thank the organisers for the invitation here to celebrate the
fiftieth anniversary of IFT, UNESP, and
 for the splendid organisation and technical facilities. I would also like to
 congratulate
them for their coup in organising as  opening event the victory 
 for the Brazilian football team in the
World Cup.

This is a conference devoted to two related topics, 
 integrability and  duality, with
the emphasis on the former, judging by the programme.
I shall start by trying to explain how, in my own mind, these ideas seem to be related before briefly mentioning some of my recent work on Lorentzian
 Kac-Moody  algebras which was  motivated by these synergies.

Most people here are connected to the theoretical physics community and  appreciate very well that the search for a comprehensive theory of interacting particles is a powerful motivation.

Thirty five years ago this motivation led me into the embryonic string theory and I soon joined those impressed by the extent that novel algebraic structures underlay the consistency of this theory  of particles and their interactions. I
am thinking of the Virasoro algebra  of gauge conditions, the vertex operators ``creating'' particles and the ``no ghost theorem'', all involving
infinite dimensional structures. This theory eventually spread worldwide
and has continued its relentless development until now, albeit in fits and starts.

Thirty years ago, Nielsen and Olesen sought to realise
 the string as a soliton in a conventional, local field theory 
and found that the appropriate  scenario was a spontaneously
broken gauge theory [1]. Their string was made of magnetic lines of force and so could only end
on a magnetic charge. The search for suitable end points prompted 't Hooft and Polyakov [2] to discover that slightly different (and more realistic) versions of  spontaneously
broken gauge theories could produce solitons mimicing particles with finitely
extended but localised structure. Stability against decay was guaranteed by a topological quantum number which could be interpreted as magnetic charge. So unified gauge theories, being spontaneously broken, can easily exhibit the tendency to spawn magnetic monopoles, those objects whose quantum consistency
was first characterised by Dirac in 1931 [3]. He found that, contrary to
 previous belief, the presence of
magnetic charge could be consistent with quantum mechanics. The proviso was
that  electric charges had to be  quantised, perfectly in line with what had been observed  and had been  hitherto inexplicable. It is ironic that 
one of the original motivations for unified theories was to find an alternative explanation of the quantisation of electric charge.

Since these solitonic monopoles have to be interpreted as particles there must be a quantum field theoretic interpretation of them rather than just the original
classical field theory one. This means there must be a quantum field operator
``creating''  them and that it ought to obey equations of motion following from some action principle. There is quite an old prototype for this, found by Skyrme in 1961, when he considered the solitons of sine-Gordon theory, non-linear equations for a real scalar field in space-time of two dimensions [4]. The quantum version of the soliton was created by an exponential of the scalar field, later called a vertex operator because of its structural resemblance to the previously mentioned
vertex operators found (later) in string theory.
 Thus the quantum soliton theory
too possesses rich and unexpected algebraic structure.

In the sine-Gordon example the soliton quantum field operator obeyed equations
of motion quite unlike the original sine-Gordon one. But when the gauge group is $SO(3)$ in the monopole situation, the known properties of the spectrum of particle states suggest that the putative monopole quantum field operator
ought to satisfy equations of motion rather similar to that of electrically charged gauge fields [5]. Some consequences of this idea could be successfully checked without constructing the analogue of the vertex operator.
 Since that construction remains unrealised even now, the idea remains
 a conjecture that will be referred to
as the electromagnetic duality conjecture for the following reasons. 

The interesting new feature of these gauge theory solitons in space-time $R^{3,1}$ is the similarity, or even symmetry, between the electrically and magnetically charged particle states, quantum excitations and quantum solitons respectively according to the initial formulation.
This is evidently related to an old idea of electromagnetic duality referring to the similarity between the behaviour of the electric and magnetic fields revealed by Maxwell's discovery of his equations. In fact it can be regarded as a quantum version of it.

Because of the difficulty in proceeding further with these ideas 
referring to space-time $R^{3,1}$ it seemed sensible  to study further
relativistically invariant soliton theories in two-dimensional space-time, $R^{1,1}$ and possible underlying algebraic structures. Such theories are definitely
more tractable and it was discovered that part of the reason for this was that
they constituted
special sorts of deformation of conformally invariant theories (familiar from the world-sheet formulation of string theories) according to an important insight of Zamolodchikov [6]. The fact that the conformal group of $R^{1,1}$ has infinite dimension means that the related integrable field theories inherit an infinite number of local conservation laws
that can be regarded as extra sorts of momenta carrying spins of magnitude
 higher
 than unity.
$$[T,P_k]=kP_k, \quad [P_k,P_m]=0;\qquad k,m\quad\subset\quad\hbox{exponents}.$$
 With this interpretation these generate an infinite extension of the space-time symmetry algebra (the Poincar\'e algebra with boost $T$ and light-cone
 momenta $P_{\pm1}$) and provide the basis of many of 
the integrability properties and special properties of the soliton solutions.
In the case of the affine Toda field theories this extended space-time algebra
is actually the principal Heisenberg subalgebra of  the centre-free version
of the underlying affine (untwisted) Kac-Moody algebra $\hat g$. The spectrum
of exponents is then given by integers equal to what are known as the exponents of the Lie algebra $g$, modulo its Coxeter number. Thus for the sine-Gordon
equation, for which $g=su(2)$, the exponents are given by all odd integers, 
positive or negative.

In the meantime there were advances in the understanding of monopole solitons
in $R^{3,1}$, some due to physicists and some due to pure mathematicians.
One was the role of supersymmetry (which came for free) in promoting certain classical properties to quantum validity [7]. Supersymmetry constitutes an enlargement of space-time symmetry as did the extra conservation laws in $R^{1,1}$
but is,
of course, different. Likewise the monopoles that have been dubbed solitons
do not possess such distinctive properties as those in $R^{1,1}$ but
they do possess a sort of weaker  integrability property that made it possible
to establish the existence of solutions describing any configuration of $N$
 static monopole solitons of like magnetic charge [8]. For each number $N$, the moduli space of solutions constituted a $4N$-dimensional manifold endowed with a metric with a hyper-K\"ahler structure [9]. This development 
culminated in 1994 when A Sen
saw how to synthesise these facts [10] to confirm the validity of some quite intricate
predictions of the electromagnetic duality conjecture. This greatly
enhanced its credibility and many of you have already heard me explain this at greater length [11] and will be aware of the many subsequent developments.

Now I want to mention a parallel and independent development in which ideas
of electromagnetic duality and the enlargement of space-time symmetry also played a role. This arose in the context of what are called ``extended supergravity
theories''. These are gravitational theories in curved space-time (${\cal M}_{n,1}$) with such a large degree of local supersymmetry that the superpartners of the graviton include both gauge particles and scalar particles. Coming with the high degree of supersymmetry is a compact symmetry group called the ${\cal R}$-symmetry group and denoted $G(\cal R)$.
It was found that $G(\cal R)$ could be enlarged to a noncompact group, $G(\hbox{Duality})$ by adding
symmetries of the equations of motion and the energy momentum tensor. This was done by including duality transformations of the gauge fields (which are abelian). The scalar fields partnered with the graviton then served as coordinates on the non-compact Riemannian symmetric space that is formed by the non-compact group $G(\hbox{Duality})$ quotiented by 
the ${\cal R}$-symmetry group (its maximal compact subgroup).

It is this  non-compact symmetry group that has become a centre of attention
because of its connection with the duality that is becoming an ever more important organising principle, particularly in superstring theory.
The first remarkable result [12] concerned  the so-called ``maximal'' supergravity theories, namely those with the highest degree of supersymmetry consistent with 
the graviton being the particle carrying the highest spin in space-time of given dimension.
Then calculation for $n$ running from $2$ to $10$ shows that the non-compact symmetry group belongs to the exceptional series of Lie algebras, with $E_{10-n}$ occurring in $R^{n,1}$ (or in a curved space-time 
$ {\cal M}_{n,1}$ with metric of the same signature). Thus
$$G({\cal R})\subset G(\hbox{Duality})\equiv E_{10-n}\quad\hbox{on}\quad{\cal M}_{n,1}.$$

The exceptional Lie algebras $E_r$ constitute a nested sequence of subalgebras:$$E_3(\equiv A_1\oplus A_2)\subset E_4(\equiv A_4)\subset E_5(\equiv D_5)\subset E_6
\subset E_7\subset E_8\subset E_9(\equiv\hat E_8)\subset E_{10}\subset E_{11}\dots $$
The Dynkin diagram of $E_r$ consists of $r$ nodes linked by single bonds in such a way that there is a central node to which are attached three legs containing $1$, $2$ and $r-4$ nodes respectively. The above sequence is obtained by lengthening or shortening the third of these legs.

 The first step had been the  discovery [13] by  Cremmer and Julia of non-compact $E_7$ in $R^{3,1}$ (with ${\cal R}$-symmetry group $SU(8)$). In $R^{10,1}$ the noncompact symmetry disappears and one can expect to find the simplest possible supergravity theory, as Nahm predicted [14]. Indeed Cremmer, Julia and Scherk
found that the only field besides those of the graviton and gravitino was an abelian
3-form field potential that enjoys a Chern Simons self-interaction
and the ability to couple naturally and geometrically to
 the world 3-volume of a  membrane [15].

Also, as Nahm had predicted, this result means that there are no supergravity theories with $n>10$.
At the other extreme, in $R^{1,1}$ the algebra $E_9$ is not a finite-dimensional Lie algebra but an affine Kac-Moody algebra, the affinisation of $E_8$ 
(often denoted $\hat E_8$)
exactly in line with the sorts of infinite algebraic structure
 already met in  the context of conformal and integrable field theories
 in $R^{1,1}$. The analogous result for Einstein's theory viewed in $R^{1,1}$
involves an infinite dimensional structure related to $\hat A_1={ S\hat U(2)}$ 
and  known to relativists as the Geroch group. 

 The most extreme case of all,  $R^{0,1}$, is particularly intriguing.
With no space coordinate the theory must be some sort of particle dynamics
related to $E_{10}$ if the pattern is to persist. $E_{10}$ itself is an interesting Lie algebra,
being neither of finite-dimensional nor affine Kac-Moody type but in fact
the largest (simply-laced) hyperbolic algebra, as will be explained.
Its Cartan matrix has ten rows and columns and determinant equal to $-1$.
There is good  reason to think that the particle mechanics refers to massless motions
within the $10$-dimensional Weyl chamber of $E_{10}$, confined to it
 by reflections off the walls and displaying chaotic behaviour, maybe a limiting case of $E_{10}$ Toda mechanics [16].

There is an apparently heretical proposal, but nevertheless one that is very intriguing, put forward by Peter West [17]. He has arguments to support the idea that
M-theory, the putative  theory of quantum super-membranes on ${\cal M}_{10,1}$, actually does possess a hidden $E_{10}$ symmetry,
exactly the symmetry realised only in $R^{1,1}$ according to the previous ideas.
It contains the conformal group of $R^{10,1}$, namely $SO(11,2)$ realised in some non-linear way. Of course these conformal groups did not play any role in
the structure previously discussed. Actually he also argues that the symmetry structure may be even larger, at least $E_{11}$. There may also be
 an analogous symmetry algebra for the bosonic string theory in $R^{25,1}$, which he calls $K_{27}$.

Neither of these can be hyperbolic as $10$ is the
 highest possible rank for them but they are both Lorentzian.
I now explain what this means and use West's conjecture as a motivation for
investigating properties of a class of these algebras.

To do this I shall briefly recap the notions underlying what are called
 Kac-Moody algebras as explained in more length in the classic book of Victor 
Kac [18]. 
There is a connected Dynkin diagram encoding the Cartan matrix $K$. The latter is assumed, for simplicity, to be symmetric. It has $r$ rows and columns, entries equal to $2$ on the diagonal and  equal to integers less than or equal to zero off the diagonal. The simple roots $\alpha_1$, $\alpha_2\dots \alpha_r$, are a set of $r$ linearly independent vectors whose scalar products are given by the elements of the Cartan matrix:
$$\alpha_i.\alpha_j=K_{ij}.$$
From this data may be deduced, in principle, the system of all the roots, which may very well be infinite. Then it is possible to specify, in principle,
commutation relations for the whole Kac-Moody algebra, denoted $g(K)$ which is well defined according to a theorem of Serre.

Out of the infinitely many possiblities for $K$ I shall consider only the following three classes:

\noindent {\bf F} : $K$ is positive definite: then $g(K)$ is a finite dimensional Lie algebra.

\noindent {\bf A} : $K$ is positive semi-definite: then $g(K)$ is an affine Kac-Moody algebra.

\noindent {\bf L} : $K$ is nonsingular with one eigenvalue negative so that 
 the remainder are positive: then $g(K)$ is said to be a Lorentzian
algebra.

Of course, classes {\bf F} and {\bf A} are completely classified, as is a subclass of {\bf L}, the hyperbolic algebras. The paper [19] by M Gaberdiel, P West and myself discusses a particularly amenable class that includes all of {\bf F}, {\bf A} and a subclass of {\bf L} that includes the hyperbolics,
 as well as the aforementioned $E_{11}$ and $K_{27}$ thought to be relevant to M-theory 
and bosonic string theory. 

This class is characterised by possessing a connected Dynkin diagram in which it is possible to choose a node, called the central node, whose deletion breaks
the original diagram into a finite number of disconnected pieces which are each of type {\bf F} or {\bf A} with no more than one piece of type {\bf A}.

It is possible to consider the Cartan Matrix $K$ for such a diagram as being bordered by the row and column corresponding to the central node. This feature can be exploited to provide a formula  for the determinant, det$K$, of the Cartan matrix of this diagram in terms of the Cartan matrices of the pieces. Similar constructions are given for the simple roots and the fundamental weights $\lambda_1,\dots \lambda_r$ defined as forming the  basis reciprocal to that provided by the simple roots:
$$\lambda_i.\alpha_j=\delta_{ij}\quad \hbox{so} \quad \lambda_i.\lambda_j=(K^{-1})_{ij}.$$

Then it is proven that $K$ is of type {\bf F}, {\bf A} or {\bf L} according as
det$K$ is positive, zero or negative, as $K$ cannot have more than one negative eigenvalue when it has the structure described.

A useful and much simplified special case occurs when each disconnected
 component is linked to the central node by just one link.
Then if the disconnected pieces have Cartan matrices $K(\hat g_0)$, $K(g_1)$,
$K(g_2)\dots K(g_n)$
$$\hbox{det}K=-\hbox{det}K(g_0)\hbox{det}K(g_1)\dots\hbox{det}K(g_n)$$
and so is negative. Thus $K$ is Lorentzian.
The factors on the right-hand side are all determinants of Cartan matrices of 
type {\bf F}. It is known to follow from the above definition of fundamental weights that then 
$$\hbox{det}K(g)=|Z(g)|$$
where $Z(g)$ is the centre of the simply connected Lie group with finite 
dimensional Lie algebra $g$. These quantities can then be easily evaluated by application of the previous identity. For example it easily follows that for any $r\geq3$
$$\hbox{det}K(E_{r})=9-r, \quad r\geq 3.$$
So $E_r$ is type {\bf F}, {\bf A} or {\bf L} according as $r\leq8,r=9$ or
 $r\geq10$ respectively. In particular det$K(E_{9})=0$,  det$K(E_{10})=-1$
 and det$K(E_{11})=-2$.

$E_{11}$ is an example of a type of Lorentzian algebra within the special 
subclass that may be called \lq\lq very extended''. For such algebras the central node
has two incident lines, one linking it to the affine node of the Dynkin diagram of $\hat g_0$ and the other to the single node of the Dynkin diagram for $A_1$.
 So
$$\hbox{det}K=-\hbox{det}K(g_0)\hbox{det}K(A_1)=-2\,\hbox{det}K(g_0).$$
$E_{11}$ and $K_{27}$ are both of this type with $g_0$ equal to $E_8$ and $D_{24}$ respectively. Hence det$K(K_{27})=-8$.

The occurrence of even unimodular lattices is often related to interesting situations in mathematics and physics (for example, 
in the theory of modular forms, in heterotic string theory etc). The root lattice of a simply laced Kac-Moody algebra
 constitutes such a lattice if det$K=\pm1$. When the  above formula applies this is satisfied if and only if $g_0=g_1=\dots g_n=E_8$, since $E_8$ is the only simply laced Lie algebra whose Cartan matrix has determinant equal to unity. 
Since the manner of linking the central node to each $E_8$ Dynkin diagram is unspecified this represents a large number of possibilities and there are even more of this type if the more general formula for det$K$ (not written down here) is considered. All these examples yield lattices of dimension $8n+10$. Thus there are very many different Lorentzian Kac-Moody algebras with the same
 root lattice, namely $II_{8(n+1),1}$, which is the unique even, unimodular, Lorentzian lattice of that dimension.

I now say something about the geometrical arrangement of the simple roots and fundamental weights. For type {\bf F} all these lie in a Euclidean space. Each simple root has length $\sqrt2$. Simple roots of type {\bf L} also have length $\sqrt2$. But they lie in a space with \lq\lq mainly positive'' Lorentzian metric and so are space-like, that is, lying outside both light cones. For the fundamental weights the situation is unclear. The distinctive feature of  hyperbolic algebras is that each fundamental weight is either time or light like. In fact they all live in or on the same light cone. So the Weyl vector,
$$\rho=\sum_{j=1}^r\lambda_j,$$
whose scalar product with each of the simple roots equals unity, lies strictly
inside this  common light cone. 

For type {\bf F} algebras this Weyl vector has an interesting relation,
 discovered by the mathematician Kostant [20], with the exponents of $g$ mentioned in connection with the Lorentz spins of the local conserved charges in integrable field theories in $R^{1,1}$. $\rho$ can be expressed as a linear combination of simple roots with strictly positive coefficients $p_i$,
$\rho=\sum_ip_i\alpha_i$, and as a consequence it is possible to construct a
{\it  real}
$SO(3)$ subalgebra of $g$ (the principal $SO(3)$ subalgebra):
$$ [T^3,T^{\pm}]=\pm T^{\pm}, \quad [T^+,T^-]=T^3,$$
by$$T^3=\rho.H,\quad T^+=\sum_{j=1}^r\sqrt{p_j}E^{\alpha_j},\quad T^-=(T^+)^{\dagger}.$$
When the adjoint representation of $g$ is decomposed into irreducible representations with  respect to this subalgebra exactly $r=$ rank$\,g$ multiplets are obtained. Their spins are precisely the exponents of $g$.
 The mathematical interest of this is that there is a beautiful formula for computing the Betti numbers of the group manifold
from these exponents.

It is natural to ask whether anything similar can be done for algebras of type 
{\bf L}. The answer is that for the class considered there is no such {\it real} $SO(3)$ subalgebra but that there is sometimes a {\it real} $SO(2,1)$ subalgebra.
For example this is so for all the hyperbolics and for $E_{11}$ but not for
 $K_{27}$.

The point is that the coefficient of the $j$'th simple root in
the expansion of the Weyl vector has the form $p_j=\rho.\lambda_j$ and so 
is always negative
in the hyperbolic case as it is a scalar product of two vectors in the same light cone. This coefficient is also always negative for the central node of the class of Lorentzian algebras considered. The above construction of the {\it real} three dimensional subalgebra only works if the coefficients $p_j$ in the expansion of the Weyl vector all have the same sign, and yields $SO(3)$ if the common sign is positive and $SO(2,1)$ if it is negative. As $\rho^2=\sum_0^rp_j$, the Weyl vector $\rho$ is time-like in the latter case. For some subclasses of Lorentzian algebra this condition is quite restrictive. Thus it requires
the rank of \lq\lq overextended'' algebras to be less than $26$, an
 intriguing result in view of the importance of that dimension in
 bosonic string theory [21]. These are algebras whose diagram consists of an affine diagram
whose affine node is linked to the central node.  But other subclasses have now been found with indefinitely high rank and time-like Weyl vector [19].

Since the group is not compact, the  unitary representations of $SO(2,1)$ are of infinite dimension. It was shown by Nicolai and myself [22] that for the hyperbolics,
once the trivial three dimensional non-unitary representation is taken out, the
the adjoint can be decomposed into unitary representations of the $SO(2,1)$.
Only $r-1$ of these possess states with $T^3=0$ and their ``spins'' 
can be used to define  $r-1$ analogues of the exponents in addition to the value $1$ associated with the triplet. These are complex numbers 
with $-1/2$ as real part and examples have been calculated. But unfortunately it is not clear what the significance of these numbers is.

Let  me conclude  by saying that there remain many open
 questions in all directions
concerning both integrability and duality but that I believe there is some real advantage in considering the topics together.
\vskip20pt
\def\ni{\noindent}

\ni [1] H.~B.~Nielsen and P.~Olesen,
``Vortex-Line Models For Dual Strings,''
Nucl.\ Phys.\ B {\bf 61} (1973) 45.

\ni [2] G.~'t Hooft,
``Magnetic Monopoles In Unified Gauge Theories,''
Nucl.\ Phys.\ B {\bf 79} (1974) 276.

\ni A.~M.~Polyakov, ``Particle Spectrum in Quantum Field Theory''
JETP Lett. {\bf 20} (1974) 194-195.

\ni [3] P.~A.~M.~Dirac,
``Quantised Singularities In The Electromagnetic Field,''
Proc.\ Roy.\ Soc.\ Lond.\ A {\bf 133} (1931) 60.

\ni [4] T.~H.~R.~Skyrme,
``Particle States Of A Quantized Meson Field,''
Proc.\ Roy.\ Soc.\ Lond.\ A {\bf 262} (1961) 237.

\ni [5] C.~Montonen and D.~I.~Olive,
``Magnetic Monopoles As Gauge Particles?,''
Phys.\ Lett.\ B {\bf 72} (1977) 117.

\ni [6] A.~B.~Zamolodchikov, ``Integrable Field Theory from Conformal Field theory''
Advanced Studies in Pure Mathematics {\bf 19} (1989) 642-674.

\ni [7] A.~D'Adda, R.~Horsley and P.~Di Vecchia,
``Supersymmetric Magnetic Monopoles And Dyons,''
Phys.\ Lett.\ B {\bf 76} (1978) 298,

\ni E.~Witten and D.~I.~Olive,
``Supersymmetry Algebras That Include Topological Charges,''
Phys.\ Lett.\ B {\bf 78} (1978) 97,

\ni H.~Osborn,
``Topological Charges For N=4 Supersymmetric Gauge Theories And
 Monopoles Of Spin 1,''
Phys.\ Lett.\ B {\bf 83} (1979) 321.

\ni [8] N.~S.~Manton,
``A Remark On The Scattering Of BPS Monopoles,''
Phys.\ Lett.\ B {\bf 110} (1982) 54.

\ni [9] M.~F.~Atiyah and N.~J.~Hitchin,
``Low-Energy Scattering Of Nonabelian Monopoles,''
Phys.\ Lett.\ A {\bf 107} (1985) 21.

\ni [10] A.~Sen,
``Dyon - monopole bound states, selfdual harmonic forms on the multi - monopole moduli space, and SL(2,Z) invariance in string theory,''
Phys.\ Lett.\ B {\bf 329} (1994) 217
[arXiv:hep-th/9402032].

\ni [11] D.~I.~Olive,
``Exact electromagnetic duality,''
Nucl.\ Phys.\ Proc.\ Suppl.\  {\bf 45A} (1996) 88
[Nucl.\ Phys.\ Proc.\ Suppl.\  {\bf 46} (1996) 1]
[arXiv:hep-th/9508089].

\ni [12] B.~Julia, LPTENS 80/1, Invited paper presented at Nuffield Gravity Workshop, Cambidge, 1980.

\ni [13] E.~Cremmer and B.~Julia,
``The SO(8) Supergravity,''
Nucl.\ Phys.\ B {\bf 159} (1979) 141.

\ni [14] W.~Nahm,
``Supersymmetries And Their Representations,''
Nucl.\ Phys.\ B {\bf 135} (1978) 149.

\ni [15] E.~Cremmer, B.~Julia and J.~Scherk,
``Supergravity Theory In 11 Dimensions,''
Phys.\ Lett.\ B {\bf 76} (1978) 409.

\ni [16] T.~Damour and M.~Henneaux,
Phys.\ Rev.\ Lett.\  {\bf 86} (2001) 4749
[arXiv:hep-th/0012172],

\ni T.~Damour, M.~Henneaux, B.~Julia and H.~Nicolai,
``Hyperbolic Kac-Moody algebras and chaos in Kaluza-Klein models,''
Phys.\ Lett.\ B {\bf 509} (2001) 323
[arXiv:hep-th/0103094].

\ni [17] P.~C.~West,
``Hidden superconformal symmetry in M theory,''
JHEP {\bf 0008} (2000) 007
[arXiv:hep-th/0005270],

\ni P.~C.~West,
``E(11) and M theory,''
Class.\ Quant.\ Grav.\  {\bf 18} (2001) 4443
[arXiv:hep-th/0104081],

\ni I.~Schnakenburg and P.~C.~West,
``Kac-Moody symmetries of IIB supergravity,''
Phys.\ Lett.\ B {\bf 517} (2001) 421
[arXiv:hep-th/0107181].

\ni [18] V.~G.~Kac, ``Infinite Dimensional Lie Algebras'', Cambridge University Press (1990) [3rd  edition].

\ni [19] M.~R.~Gaberdiel, D.~I.~Olive and P.~C.~West,
``A class of Lorentzian Kac-Moody algebras,''
arXiv:hep-th/0205068.

\ni [20] B.~Kostant, Am. J. Math. {\bf 81} (1959) 973.

\ni [21] P.~Goddard and D.~I.~Olive, ``Algebras, Lattices and Strings'', in
``Vertex Operators in Mathematics and Physics'' MSRI Publications $\sharp 3$,
 Springer (1984) 51.

\ni [22] H.~Nicolai and D.~I.~Olive,
``The principal SO(1,2) subalgebra of a hyperbolic Kac Moody algebra,''
Lett.\ Math.\ Phys.\  {\bf 58} (2001) 141
[arXiv:hep-th/0107146].

\end